\documentclass[11pt,twoside]{article}  
\usepackage{apn3conf}

\begin{document}   

%
%
%
%

\title{Kinematics and Proper Motion of the Ansae in NGC\,7009}

%
%
%

\author{Rodrigo Fern\'andez, Hugo E. Schwarz, Hektor Monteiro}
\affil{Cerro Tololo Inter-American Observatory, NOAO-AURA, Casilla 603, 
La Serena, Chile}

%
%

\contact{Rodrigo Fern\'andez}
\email{rafernan@puc.cl}

%
%
%
%
%

\paindex{Fern\'andez, R.}
\aindex{Schwarz, H.}     
\aindex{Monteiro, H.}

%
%

\authormark{Fern\'andez, Schwarz \& Monteiro}

%
%



\begin{abstract}          
We have measured the expansion velocities and proper motion of the
ansae in NGC\,7009 using high dispersion echelle spectra and archive
narrow band HST images. Assuming that the ansae are moving at equal
and opposite velocities from the central star we obtain an average
system radial velocity of $-54 \pm 2$ km/s, the eastern ansa
approaching and the western ansa receding at $v_r=5.5 \pm 1$ km/s
relative to this value. Only the proper motion of the eastern ansa
could be measured, leading to $2.8 \pm 0.8$ arcsec/century, or
$v_t=(130 \pm 40)d$ km/s, where $d$ is the distance to the nebula in
$kpc$. Additionally, the electron temperature and density for each
ansa was measured using line intensity ratios.  The results are $T_e
\sim 9000$ K and $n_e \sim 2000$ $cm^{-3}$ for both ansae within the errors.
\end{abstract}

%
%

\section{Introduction}

NGC\,7009 is a good example of a planetary nebula with
FLIERs\footnote{Fast, Low Ionization Emission Regions (Balick et
al. 1994).}  The nebula shows two pairs of condensations along the
major symmetry axis. The outer pair has been called
\emph{ansae} (=handles, Aller 1941). The condensations show strong 
emission in low ionization lines and weak emission in high excitation
ones. Kinematic studies by Reay \& Atherton~(1985) and Balick, Preston
\& Icke~(1987) have shown that the ansae are moving very near the
plane of the sky at velocities $\sim 10^2$ km/s with respect to the
central star.  Here we aim to better determine the kinematics of the
ansae in NGC\,7009, measuring radial velocities from high dispersion
echelle spectra and proper motions from archive HST images. As a
by-product, the electron temperatures and densities are measured from
line intensity ratios.

\section{Observations and Data Reduction}

The data consist of a set of echelle spectra and two
sets of public HST images taken in 1996 and 2001.  The spectra were
obtained during the night of July\,29, 2002, with the CTIO 4m echelle
spectrograph, which was optimized to cover the spectral range 410\,nm
- 720\,nm. The dimensions of the slit were $1.2\arcsec \times
6.6\arcsec$, and the mean seeing was $1.4\arcsec$. For wavelength
calibration a Th-Ar lamp was used. The flux calibration was done with
observations of the star 58\,Aql (HR\,7596).

The HST images were obtained from the HST archive. The first set was
taken on April\,28, 1996 using the WFPC2 with a [NII]\,6583 narrow band
filter (Balick et al.\,1998). The second set was taken in May\,11, 2001
also with the WFPC2 and the same filter (Palen et al.\,2002).

The echelle spectra were reduced and calibrated using standard IRAF
routines. The dispersion is 0.008\,nm\,pixel$^{-1}$, or
3.7\,km.s$^{-1}$.pixel$^{-1}$ at 650\,nm. The HST images were
retrieved from the archive using the \emph{On The Fly Calibration}
option. The calibrated images of each set were median combined to
eliminate cosmic rays.

\section{Results}

\subsection{Radial Velocities}

The lines that could be identified in the spectra are listed in Table
\ref{spectra}. The central wavelength of each line was measured by
fitting a Voigt profile. Almost every line could be identified in two
echelle orders, thus the value quoted is the average of the two values
in each order. The rest wavelengths of forbidden lines were taken from
Bowen\,(1960), while the values for the permitted lines were taken from
Moore\,(1945). Assuming symmetric expansion with respect to the central
star, the system velocity is the average of the ansae's velocities,
namely $-54\pm 2$\,km.s$^{-1}$, whereas the radial expansion velocity with
respect to the central star is $v_r = \pm 5.5$ km/s.

\begin{deluxetable}{llcc}
\tablecaption{Radial velocities for identified emission lines.}
\tablehead{Line & $\lambda$(nm) & Eastern Ansa (km.s$^{-1}$) & 
Western Ansa (km.s$^{-1}$)}
\startdata
$[$NII$]$ & 575.5 & -59.17 & -47.70\\
HeI       & 587.6 & -56.42 & -48.51\\
$[$OI$]$  & 636.4 & -62.23 & -50.44\\
$[$OI$]$  & 630.0 & -61.19 & -49.05\\
$[$NII$]$ & 654.8 & -59.33 & -48.34\\
H$\alpha$ & 656.3 & -56.69 & -48.46\\
$[$NII$]$ & 658.3 & -57.87 & -46.48\\
$[$SII$]$ & 671.6 & -57.68 & -45.34\\
$[$SII$]$ & 673.1 & -56.83 & -45.46\\
$[$AIII$]$& 713.4 & -57.60 & -48.97\\
\noalign{\smallskip}
{\bf Average} &  & -58.55 $\pm$ 1.9 & -47.82$\pm$ 1.6\\
\enddata
\label{spectra}
\end{deluxetable}

\subsection{Proper Motions}

The HST images have a shift between 1996 and 2001 of $20\arcsec$ in
the east-west direction, which makes the western ansa disappear from
the 2001 images. Therefore, only the proper motion of the eastern ansa
could be measured.  The position of the ansa was determined by
confining it in a box of width $5\arcsec$ and calculating a
centroid. Three stars were selected as reference points for angular
displacement measurements, namely the central star and two background
stars. The results for the angular displacements are listed in Table
\ref{angdisp}. A fourth reference point was determined by looking for
a point that minimized the square of the distances between the three
reference stars, weighted by the error in each position. The angular
displacement relative to this point is listed as \emph{Least
squares}. An average between the displacements with respect to each
star is also listed.  The proper motion is obtained by dividing the
corresponding angular displacement by the time difference between both
set of images, nl. 5.038170123 yrs. All displacements occur in the
same direction, outward from the central star.  Using the least
squares value for the proper motion, one gets a tangential velocity of
$v_t = (130\pm 32) d$ km/s, where $d$ is the distance to the nebula in
kpc.  Computing a weighted average of 14 values from the literature
(Acker et al.\,1992) gave 0.86kpc or $v_t =(112\pm 32) $ km/s

\begin{deluxetable}{lcc}
\tablecaption{Angular displacement and proper motion of the eastern ansa, 
using different reference points}
\tablehead{Ref. point & Ang. displacement (\arcsec) & Prop. motion 
(\arcsec/yr)}
\startdata
Central star & $0.17\pm 0.10$ & $0.03\pm 0.02$\\
Background star 1 & $0.16\pm 0.10$ & $0.03\pm 0.02$\\
Background star 2 & $0.10\pm 0.02$ & $0.020\pm 0.004$\\
Average & $0.13\pm 0.04$ & $0.026\pm 0.008$\\
Least squares & $0.14\pm 0.04$ & $0.028\pm 0.008$
\enddata
\label{angdisp}
\end{deluxetable}

\subsection{Line intensities, Electron Temperatures and Densities}

The line intensities of the [NII] and [SII] lines were measured in
order to derive the electron temperature and density in both
ansae. Voigt profiles were fitted to each line using the IRAF task
\emph{splot}. The interstellar extinction correction was made assuming
a recombination model B with $T = 10^4$\,K and $n_e =
10^4$\,cm$^{-3}$, and an extinction coefficient $c_\beta\,=\,0.26$
(Lame \& Pogge\,1996). The line intensity ratios together with electron
temperatures and densities derived from them are listed in Table
\ref{Tne}. The latter results were obtained using the formulae from
McCall\,(1984).  The errors in the ratios were calculated from the S/N
in each line.

\begin{table}[!htb]
\caption{Line intensity ratios and results obtained from them. 
$T_e$ and $n_e$ are the electron temperature
and density, respectively.}
\begin{tabular}{lccl}
\noalign{\bigskip}
\noalign{\emph{Eastern ansa}}
\noalign{\smallskip}
Ratio &                     Observed &       Corrected &        Result\\
\noalign{\smallskip}
\hline
\noalign{\smallskip}
$[$NII$]$(6548+6583)/5755 &      $111 \pm 15$ &       $103 \pm 15$ &     
   $T_e = 9000 \pm 400$ K\\
\noalign{\smallskip}
$[$SII$]$ 6716/6730 &           $0.67 \pm 0.04$ &    $0.67 \pm 0.04$ &   
   $n_e = 2600 \pm 500$ $cm^{-3}$\\
\noalign{\smallskip}
\hline
\noalign{\vspace{0.3in}}
\noalign{\emph{Western Ansa}}
\noalign{\smallskip}
Ratio &                     Observed &       Corrected &        Result\\
\noalign{\smallskip}
\hline
\noalign{\smallskip}
$[$NII$]$(6548+6583)/5755 &      $120 \pm 12$ &      $110 \pm 12$ &      
    $T_e = 8900 \pm 400$ K\\
\noalign{\smallskip}
$[$SII$]$ 6716/6730  &          $0.64 \pm 0.04$ &    $0.64 \pm 0.04$ &   
     $n_e = 1900 \pm 300$ $cm^{-3}$\\
\noalign{\smallskip}
\hline
\end{tabular}
\label{Tne}
\end{table}

\section{Comparison with previous results}

The results for the radial velocities are in good agreement with
previous determinations. If we take only the results for the [OI] 6300
line, we get a result equal within the errors to that of Reay \&
Atherton\,(1984), who measured $v_r\,=\,\pm 6.2$\,km.s$^{-1}$ with
respect to the central star using a Fabry-Perot interferometer.  The
only available measurement of proper motion is that of Liller\,(1965),
who obtained 1.6\,\arcsec/century using photographic plates, which is
almost a factor of two smaller than the result obtained here. The
electron temperatures are equal within the errors to those reported by
Balick et al.\,(1994) and Bohigas et al.\,(1994), while the electron
density is within a factor of two from the results of the same
authors.

We conclude that the ansae are moving with total velocity
$V\,=\,\sqrt{v_r^2\,+\,v_t^2} \simeq v_t\,=\,(130\pm 32)d$
km.s$^{-1}$, at an angle $i\,=\,\arctan{v_r/V}\,=\,2\pm\,1.6\,^o$ with
respect to the plane of the sky.

\acknowledgments

This work was done as part of the \emph{Programa de Investigaci\'on en
Astronom\'\i a} (PIA) for chilean students at CTIO.

%
%
%
%


\end{document}